\documentclass[journal, 12pt, draftclsnofoot, onecolumn]{IEEEtran}

\normalsize

\usepackage{cite,graphicx,epstopdf,color,amssymb,amsopn,amsthm,algorithm,algorithmic,array,multirow,eqparbox,flushend,hhline,tablefootnote,amsfonts,dsfont,threeparttable}

\usepackage[cmex10]{amsmath}

\usepackage[caption=false,font=footnotesize]{subfig}

\usepackage{fancyhdr}
\pdfoutput=1

\newcommand{\ds}[1]{\textcolor{black}{#1}}
\newcommand{\ns}[1]{\textcolor{black}{#1}}
\newcommand{\nss}[1]{\textcolor{black}{#1}}

\newtheorem{define}{Definition}
\newtheorem{proposition}{Proposition}
\newtheorem{remark}{Remark}
\newcommand{\nic}[1]{\textcolor{black}{#1}}
\newcommand{\da}[1]{\textcolor{black}{#1}}
\newcommand{\dsn}[1]{\textcolor{black}{#1}}

\begin{document}

\title{U-BeAS: A Stackelberg Game for Device-to-Device Communications}

\author{Nicole~Sawyer,~\IEEEmembership{Student Member,~IEEE,}
        and~David~Smith,~\IEEEmembership{Member,~IEEE.}
\thanks{The authors are with Data61, Canberra ACT 2601, Australia, E-mails: \{Nicole.Sawyer, David.Smith\}@data61.csiro.au.  The authors are also with the Research School of Engineering, The Australian National University, Canberra ACT 2601, Australia.}}

%

\maketitle

\begin{abstract}
User-behavior-aware communications will be a key feature of future generation cellular networks, so user experience is enhanced with greater benefit to users.  In this paper, a user-behavior-aware Stackelberg (U-BeAS) game for device-to-device (D2D) communications overlaying cellular communications is proposed.  Our proposed game provides an optimal trade-off between minimizing transmit power and maximizing packet delivery ratio (PDR), with respect to D2D user-behavior for all D2D pairs.  The Stackelberg leader, base station (BS), selects its satisfaction with the social welfare of the D2D followers by considering the reaction of the followers to the price it charges.  All followers (D2D pairs) select transmit power so as to guarantee a desired quality-of-experience (QoE), where each follower enacts one of three possible behaviors: (i) casual; (ii) intermediate; or (iii) serious. Analysis shows that there exists a sub-game perfect Stackelberg Equilibrium across all players.  Analysis and simulation demonstrates that the BS leader rapidly converges to an optimal satisfaction while guaranteeing social welfare across D2D users, and all D2D followers then rapidly converge to a Pareto-efficient outcome with respect to transmit power and PDR.
\end{abstract}


%
\IEEEpeerreviewmaketitle

\newpage
\section{Introduction}
\IEEEPARstart{I}{n} future fifth generation (5G) cellular networks, it is expected that the number of users and user demand is going to continue to increase \ds{significantly} \cite{cisco2016cisco}.  Device-to-Device (D2D) communications plays a key role in helping take the load off the cellular network by reducing the demand on the central network operator, i.e., the base station (BS) \cite{lin2014spectrum,gu2015resource,yang2013solving,fodor2012design,liu2014device}. Introducing D2D communications into the cellular network can also provide an increase in resource utilization, interference, quality-of-service (QoS), quality-of-experience (QoE), energy efficiency, spectral efficiency, system capacity, and communication reliability.  We study D2D communications as an overlay to cellular communications in order to eliminate intra-cell interference, i.e., the interference between cellular and D2D users \cite{lin2014spectrum,yin2015pricing,song2014game,liu2013resource}, as D2D pairs are allocated orthogonal spectrum.

\nic{\textit{User-behavior in cellular communications is a key aspect of future generation cellular networks}.  Considering \nic{user-behavior} alongside cellular communications, can potentially enhance user experience and \da{enable greater perceived benefit to users}.  Recently, a method incorporating \nic{user-behavior} has been proposed for cellular communications, called User-in-the-Loop (UIL).}  UIL is a closed loop system with feedback, which allows users to choose what their requirements are and whether they will allow their behavior to be controlled or influenced \cite{schoenen2012quantified,schoenen2014user,schoenen2012first,wang2015load,schoenen2013dynamic}, i.e., smart communications.  \nic{Still, user-behavior needs to be considered further for future generation cellular networks.}

\subsection{Contributions}
In this paper, we \nic{introduce} a \nic{user-behavior-aware} dynamic Stackelberg \nic{repeated} game \nic{(U-BeAS) to solve a} resource allocation problem in distributed D2D communications, with D2D overlay to cellular users.  \nic{Moreover, the proposed game aims to contribute to the user-behavior study, \ns{through an optimal trade-off between} transmit power and packet delivery ratio (PDR)\ns{, with respect to D2D user-behavior}.  Overall, U-BeAS will guarantee D2D user} \ds{satisfaction}\nic{, reduce power consumption, and increase communication reliability.  Stackelberg games can be made up of either multiple single leader-follower pairs, or a single leader with multiple followers.  We consider} \ds{a} \nic{single leader with multiple followers case, where the} BS is the Stackelberg leader, \nic{and the D2D pairs are the Stackelberg followers.  The BS} determines a price to charge the followers \nic{for using the allocated channel and obtaining} satisfaction \dsn{with} the communication cell.  The D2D pairs access the same spectrum, \nic{i.e., reuse D2D spectrum, and play a non-cooperative power control game.  The D2D pairs} aim to optimally select transmit power while guaranteeing QoS and QoE, with respect to the satisfaction price set by the BS.  Hence, the following contributions are made:
\begin{itemize}
  \item U-BeAS is a user-behavior-aware dynamic Stackelberg repeated non-cooperative game\nic{. U-BeAS} jointly optimizes user-behavior at the application layer and transmit power at the physical (PHY) layer, to increase overall system capacity, QoS, QoE, and \nic{communication reliability,} while reducing transmit power for all players, as well as guaranteeing social welfare across all D2D users.
  \item We \nic{categorize} D2D user-behavior \nic{in}to three \nic{behavior classes}: \nic{(i) casual; (ii) intermediate; and (iii) serious}.  Each behavior \nic{class} is assigned a different utility function which suits the particular behavior and resource requirements\ns{, which allows for an optimal trade-off between transmit power and PDR with respect to D2D user-behaviour for all players}.
  \item The performance metric, packet delivery ratio (PDR), is used to measure communication priority and reliability for each D2D pair.  We show the effects of priority on \nic{the proposed U-BeAS game,} by setting different minimum PDR targets for each \nic{behavior class}, e.g., \nic{casual-behavior} has a minimum target PDR of 0.90, \nic{intermediate-behavior} has a minimum target PDR of 0.94, and \nic{serious-behavior} has a minimum target PDR of 0.98.
  \item We show that our proposed game overall has a \nss{unique} Stackelberg Equilibrium, and it is a sub-game perfect equilibrium at each stage in the game, which is independent of the game history.  The leader's satisfaction first rapidly converges to an optimal outcome, then the followers rapidly converge to a Pareto-efficient transmit power and PDR outcome. The optimal outcome for the leader in terms of satisfaction helps guarantee social welfare across all D2D users.
\end{itemize}

\subsection{Related Work}
D2D communications can be implemented in licensed or unlicensed spectrum.  We study D2D communications in licensed spectrum (in-band), which is the cellular spectrum \cite{lin2014spectrum,yin2015pricing,lin2013optimal,liu2014device}.  Depending on how D2D and cellular users share the spectrum, D2D communications \nic{can potentially} introduce two-types of interference within the network: inter-cell interference and intra-cell interference.  \nic{As previously mentioned, we study D2D communications as an overlay to cellular communications, thus eliminating the possibility of intra-cell interference.}  Inter-cell interference \nic{on the other hand,} is the interference generated between D2D and cellular users from other cells \cite{lin2014spectrum,yin2015pricing,huang2014resource,liu2013resource}.  \nic{Various inter-cell interference management schemes have been implemented in order to reduce this type of interference, such as a resource allocation protocol, as in \cite{huang2014resource}, and an uplink semi-persistent scheduling resource, as in \cite{liu2013resource}.}

\nic{\nic{Game theory has been widely used to help solve optimization problems in D2D communications.  That is,} game theory is a decision making process, and allows rational players (e.g., cellular and D2D users) to select their action from an action set \cite{song2014game}.  During the game, each player aims to maximize their utility with respect to their action, to find an optimal solution and/or equilibrium in the game\cite{song2014game,huang2014resource}.  Here we will focus on non-cooperative game \nic{theory}, where no communication exists between players.  There are several types of non-coopeative games \nic{which have been used} to optimize power control for D2D communications, such as non-cooperative power control games, Stackelberg games, and auction-based games \cite{song2014game}.}  Non-cooperative power control games \nic{aim to} minimize transmit power and \nic{help mitigate/eliminate some of the} interference \nic{within the cell}, while guaranteeing QoS for all users, as in the following \cite{fodor2011distributed,fodor2013comparative,song2014game,phunchongharn2013resource}.  In \cite{sawyer2016pareto}, a non-cooperative cross-layer repeated game is proposed, which combines a non-cooperative power control game and a two armed-bandit game.  This work aims at minimizing transmit power and interference, while guaranteeing QoS and selecting optimal transmission mode for all D2D users.  Stackelberg games \nic{on the other hand, consist of a hierarchical structure with a leader and follower \cite{song2014game}.  In \cite{wang2014quality}, a Stackelberg game was applied to optimize the quality of wireless multimedia transmission in D2D communications, i.e., a quality optimized joint source selection and power control with packet prioritization.  In \cite{wang2013joint,xia2014resource,yin2013distributed}, joint power control and resource allocation (scheduling) schemes were proposed, with D2D communications as an underlay to cellular communications.  Auction based games on the other hand, consists of an auctioneer (seller) and bidders (players).  In an auction-based game, the auctioneer sets the initial price of the resource to be sold, and the players/bidders compete to win the resources up for auction \cite{song2014game}. This type of game has been used in resource allocation scenarios, where \cite{wang2015energy} proposes a joint power and resource allocation scheme to be solved using a combinatorial auction game for D2D communications.}  The aforementioned works considered interference management schemes, power control, and resource utilization, however none of the works considered joint optimization between D2D user-behaviour and resource utilization, using non-cooperative game theory.

D2D communications allows the offloading of traffic from the BS, which provides the following advantages to the network: (i) the BS can accommodate more users within the cell and increase system capacity and throughput \cite{liu2014device,song2014game,han2014user,chen2015optimal}; (ii) increasing spectral efficiency of the network \cite{liu2014device,song2014game,han2014user,chen2015optimal}; (iii) reduced latency for both cellular and D2D users \cite{liu2014device}; (iv) reducing the required power consumption of the BS, cellular users, and D2D users \cite{fodor2011distributed,fodor2013comparative,song2014game,phunchongharn2013resource,han2014user}.  As a result, the BS will have more resources to allocate to determining the cell layout, user locations within the cell, priority, and type of communications.

The BS can also measure each D2D user's satisfaction within the cell, in order to satisfy cell requirements \nic{and ensure users are satisfied}.  \nic{User satisfaction can also be expressed as} \da{user's perceived benefit,} \nic{meaning that user satisfaction is the measure of} \da{benefit that a particular user feels} \cite{han2014user}.  \nic{D2D user satisfaction can be measured in the following ways, guaranteeing QoS for all users \cite{wen2013qos},} \ds{alternatively} \nic{guaranteeing QoE for all users} \cite{schoenen2012quantified}\nic{, or by ensuring that high data rate is provided \cite{han2014user}}.  \nic{For example, if low data rate was provided to a particular user, then this would have a negative effect on user satisfaction,} \ds{w}\nic{hen compared to if high data rate was provided.}  In \cite{han2014user}, a user satisfaction based resource allocation algorithm for D2D communications was established, using an artificial fish swarm approach.  The proposed approach \nic{divides user satisfaction into three classes: (i) conversational class; (ii) streaming class; and (iii) interactive class.  Each class is assigned it own utility, specific to the required levels of satisfaction.  Overall, \cite{han2014user}} provides improved average system satisfaction and increased system throughput, as the number of D2D pairs is increased \nic{in the cell}.  Hence, the BS can determine the social welfare of the cell by summing all the D2D users satisfaction \cite{chen2015optimal,marbach2002downlink}, where social welfare is a socially optimal outcome for particular network users-of-interest \cite{dong2015socially,saraydar2002efficient}.

User-in-the-loop (UIL) is \nic{another} approach which can be used to enable, and measure, D2D user satisfaction.  UIL aims to aid user's experience by reducing traffic and delay within the network.  Within UIL users can choose their behavior \nic{or resource} requirements.  \nic{Depending on the user's} chosen behavior, incentives or penalties may be charged to the user, to ensure users maintain a certain level of QoE \cite{schoenen2012first},\cite{wang2015load}.  \ds{According to} the user's chosen behavior, the incentives provided to the user may consist of postponing their non-urgent message to another time when there is less traffic, or moving to a different location where there is less traffic \cite{schoenen2014user},\cite{schoenen2012first}.  On the other hand, the penalties provided to the user may consist of charging the user a cost, if they did not wish to postpone their message or move to a different location \cite{schoenen2014user},\cite{schoenen2012first}.  Considering user-behavior within the network, overall network spectral efficiency can be increased, along with QoE for each user \cite{schoenen2012quantified}.  Jointly optimizing user behavior and resource utilization is an important research problem that needs to be addressed for D2D communications, as outlined in \cite{andreev2015network}.  Such joint optimization, as outlined in this paper, can lead to significantly improved user experience and require less user input, while optimizing communication parameters that increase network efficiency.

The rest of this paper is organized as follows.  In Section \ref{sec:2} we outline the system model of our proposed game.  \nic{Here we also define the problem which we wish to solve, and discuss the reasoning behind the D2D user-behavior classes and what satisfaction of the BS means to the network.}  In Section \nic{\ref{sec:4}} we define \nic{a user-behavior-aware} dynamic Stackelberg repeated game (U-BeAS), \nic{and analyze the game's characteristics.  There we also outline the leader's and followers'} utility function\nic{s}, \nic{and introduce a satisfaction price that is assigned by the leader to each of the followers' utility function}.  The simulation \nic{set-up and results are shown} in Section \nic{\ref{sec:5}}.  Finally, concluding remarks are made in Section \nic{\ref{sec:6}}.

\section{System Model} \label{sec:2}
\begin{figure}[t!]
  \centering
  \includegraphics[width=0.8\columnwidth]{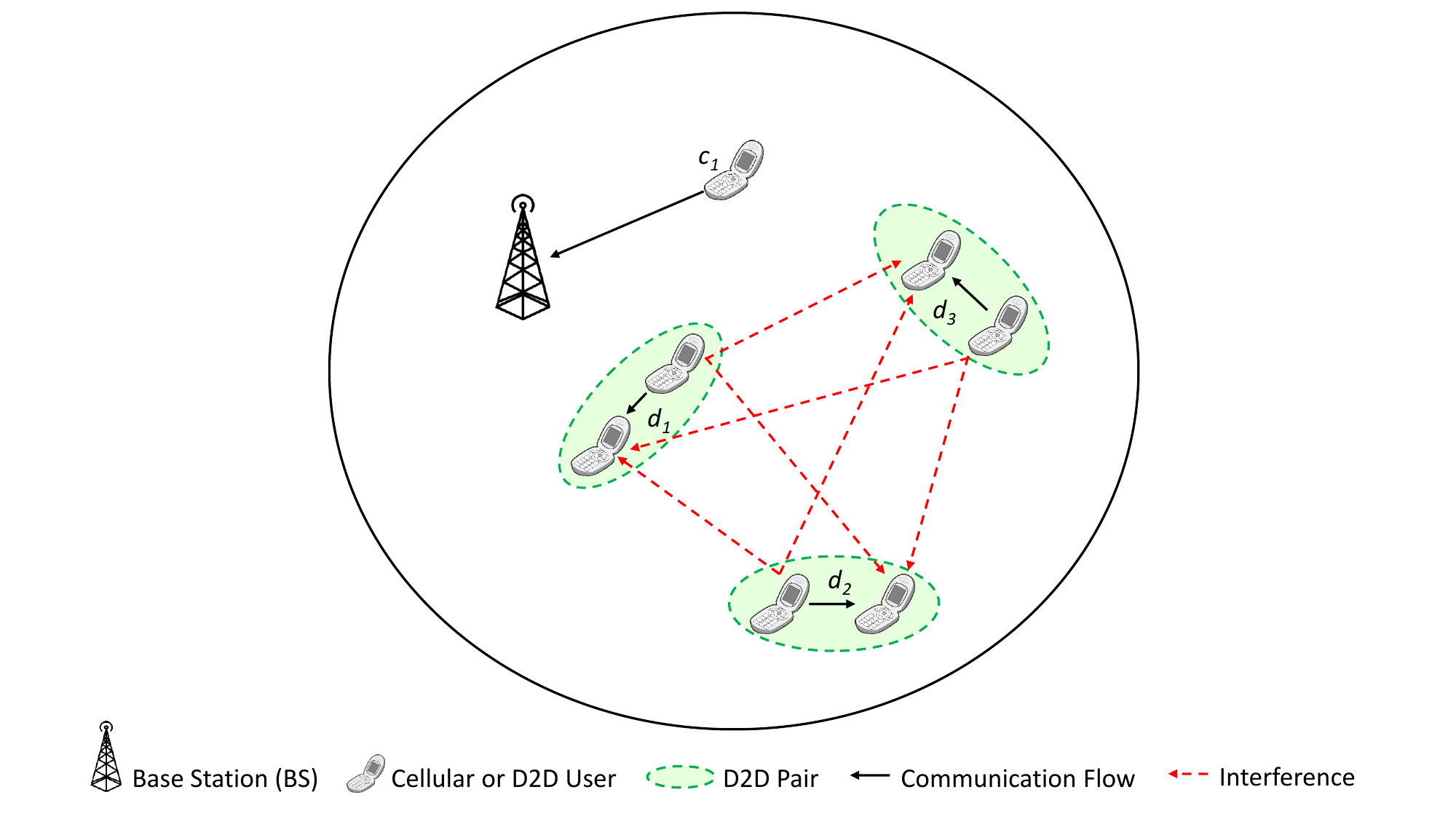}
  \caption{System topology of the U-BeAS game.}
  \label{fig:SystemModel}
\end{figure}

We implement our proposed game in a single cell for a future fifth generation (5G) cellular network, where D2D communications overlay\ds{s} the cellular network.  Future 5G cellular networks will supersede Long Term Evolution - Advanced (LTE-A) cellular networks where some forms of D2D communications can already be deployed. Within the cell there exists one base station (BS), one channel, one cellular user, and $\mathcal{M}$ D2D \nic{pairs}, \{$d_1$,$d_2$,\ldots,$d_M$\}, as illustrated in Fig. \ref{fig:SystemModel}.  This simplified scenario can be applied to a more general case with many cellular and D2D users.  For two D2D users to communicate directly to one another, both users must be within the maximum transmission range, creating a D2D transmitter-receiver pair.  \nic{The cell has been allocated one channel, which is divided into $K$ orthogonal subchannels, as in \cite{yin2015pricing,lin2013optimal,phunchongharn2013resource,wang2013joint,yin2013distributed}, where the BS allocates part of the uplink spectrum from cellular users to D2D users, and the left-over spectrum can be utilized by the cellular user.  The D2D pairs will compete against one another for the spectrum without causing interference to the cellular user \cite{yin2015pricing,lin2013optimal,song2014game,liu2013resource}, due to the overlay scenario.  (Importantly, the D2D users can still cause significant radio interference to each other, i.e., non-paired D2D transmitters act as hidden terminals to \nic{non-paired} D2D \nic{receivers} within the cell \nic{if they are utilizing the same spectrum}).  Additionally, here we consider a single cell scenario, so inter-cell interference is not considered.  Due to the spectrum allocation by the BS, both users within a D2D pair cannot transmit at the same time, as this would result in large self-interference \cite{ali2014full}.  Hence, we assume that only one user within the D2D pair transmits their message at any one time.}

Depending on spectrum allocation by the BS, interference between D2D pairs may be large.  The interference received at a D2D receiver, is the sum of the non-paired D2D transmitters sharing the same subchannel\nic{, plus \emph{additive white Gaussian noise} (AWGN) power}.  \nic{Here we considered a worst case scenario, where all D2D pairs are sharing the same spectrum}.  To \nic{ensure signal quality for all D2D pairs in the network, }we place a threshold on \emph{Signal-to-Interference-plus-Noise Ratio} (SINR), to \nic{guarantee QoS}.  The SINR at the receiver for the $i$-th D2D \nic{pair} using part of the $n$-th subchannel at time slot $t$ is given by,
    \begin{equation} \label{eq:sinrD}
        \gamma_{di,n}(t) = \frac{p_{di,n}(t)|g_{dii}|^2}{\sum^M_{\substack{j=1\\j\neq i}}p_{dj,n}(t)|g_{dji}|^2+N_{0,d}},
    \end{equation}
    \begin{equation}
        \gamma_{di,n} \geq \bar{\gamma}_d,
    \end{equation}
where $p_{di,n}(t)$ is transmit power of the $i$-th D2D \nic{pair} in subchannel $n$ at time $t$; $p_{dj,n}(t)$ is the interfering transmit power of the $j$-th D2D \nic{pair}, who is sharing the same subchannel as the $i$-th D2D user; $g_{dii}$ is the channel gain \nic{of the $i$-th} D2D pair; $g_{dji}$ is the interfering channel gain between the $i$-th D2D pair receiver and $j$-th D2D \nic{pair} transmitter, in the same subchannel; $N_{0,d}$ is the \nic{AWGN} power for D2D users; and $\bar{\gamma}_d$ is the target SINR for D2D users. 

\subsection{Channel Model}
The channel model is based on free space path loss and small-scale fading (i.e., Rayleigh fading), as follows,
    \begin{equation}
        g_{dij} = A_{PL}A_{SSF}\bigg(\frac{d_0}{d_{ij}}\bigg)^{\frac{\alpha}{2}},
    \end{equation}
where $A_{PL}$ is the free space path loss channel attenuation; $A_{SSF}$ is the small-scale fading (i.e., Rayleigh fading) channel attenuation, calculated using Jakes model; $d_0$ is the reference distance between \nic{a transmitter and receiver}; $d_{ij}$ is the distance between the $i$-th transmitter and the $j$-th receiver; $\alpha$ is the path loss exponent.

\subsection{Performance Metric - Packet Delivery Ratio}\label{sec:2.1}
The performance metric we \da{choose} to measure our proposed game \da{with} is \emph{packet delivery ratio} (PDR).  PDR is a compressed exponential function of inverse SINR \cite{smith2014multi}, as follows,
    \begin{equation}\label{eq:pdri}
        pdr_{di} = \exp\bigg(-\bigg(\frac{1}{\gamma_{di} a_c}\bigg)^{b_c} \bigg),
    \end{equation}
where $\gamma_{di}$ is the SINR of the $i$-th D2D user; $a_c$ and $b_c$ are constants which depend on packet size, type of modulation, and coding scheme.  We can rearrange \eqref{eq:pdri} into a simpler form as follows,
    \begin{equation}\label{eq:pdr}
        pdr_{di} = \exp\big(a{\gamma_{di}}^b\big),
    \end{equation}
    \begin{equation*}
        a = -\bigg(\frac{1}{a_c}\bigg)^{b_c},~b = -b_c.
    \end{equation*}

\nic{To ensure acceptable PDR is achieved} \da{we seek to meet or exceed a target PDR, such that} \nic{$pdr_{di} \geq pdr_{tgt}$, where $pdr_{tgt}$ is target PDR} \ns{and is defined} \da{in terms of target SINR $\bar{\gamma}_d$} \ns{as, $pdr_{tgt}=\exp(a(\bar{\gamma}_d)^b)$}. \da{(Conversely, thus target SINR, $\bar{\gamma}_d=\big[\frac{\ln(pdr_{tgt})}{a}\big]^{\frac{1}{b}}$)}.  \nic{Ideally, we want users to have} \da{very good perceived benefit} \nic{throughout the} cell, which includes \nic{achieving PDR above target PDR, i.e., guaranteeing QoS and QoE} \da{and maintaining desired communications reliability}\nic{.  PDR can be used to also} \da{prioritize communications}\nic{, by assigning particular users a higher target PDR than others, this means that higher target PDR entails higher communications priority and} \da{greater} \nic{reliability.}

\begin{figure}[!t]
        \centering
        \includegraphics[width=0.63\columnwidth]{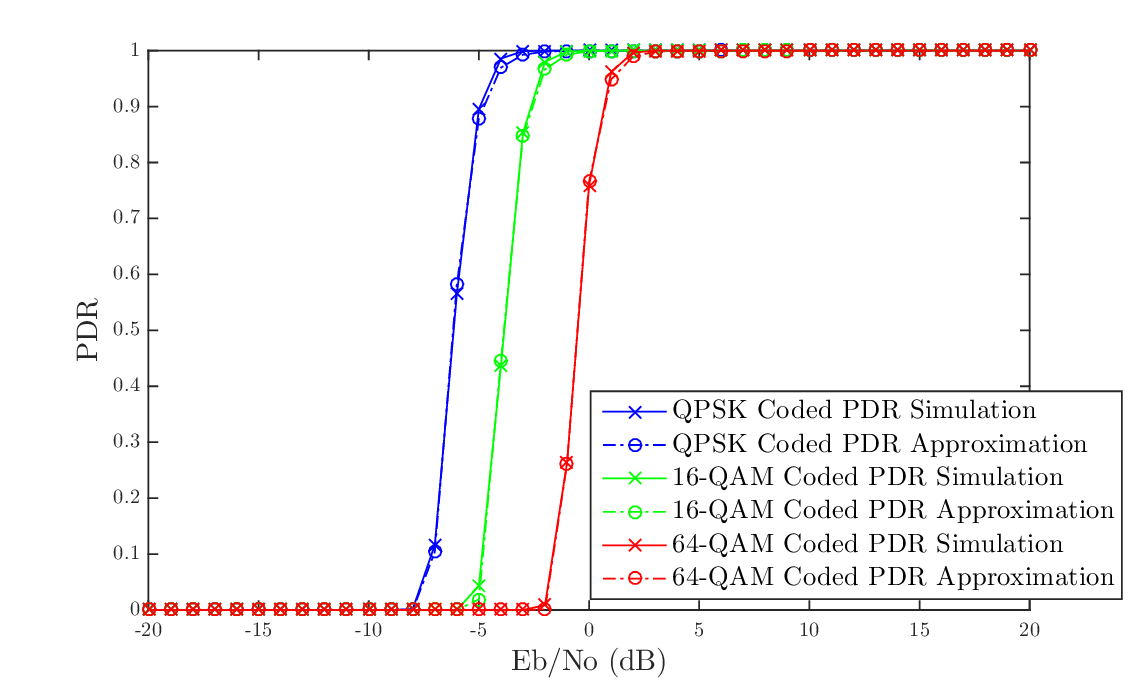}
        \caption{Comparing simulation and approximation results for PDR vs SINR.}
        \label{fig:PDRvSINR}
    \end{figure}

    \begin{table} [!t]
        \begin{center}
        \caption{Estimated parameters, $a_c$ and $b_c$, from \eqref{eq:pdri}}\label{tab:pdrparam}
            \begin{tabular}{ | l | l | l | l| l | l |}
                \hline
                \textbf{Modulation} & \textbf{Coding Gain} & $a_c$ & $b_c$ &   $a$   &  $b$   \\ \hline
                QPSK       &  13.75 dB   & 2.331 & 6.355 & -0.0001 & -6.22  \\ \hline
                16-QAM     &  15.75 dB   & 1.383 & 6.565 & -0.0014 & -6.88  \\ \hline
                64-QAM     &  17 dB      & 0.762 & 7.014 & -0.2669 & -7.021 \\
                \hline
            \end{tabular}
        \end{center}
    \end{table}

According to the 3GPP standards, the modulation schemes employed \nic{will be} QPSK, 16-QAM, and 64-QAM, with a turbo code rate 1/3 \cite{zyren2007overview}.  We are able to solve \emph{Bit Error Rate} (BER) with respect to SINR $\big(\frac{E_b}{N_o}\big)$ for the three modulations, with no coding and with coding for the same packet length of 1024 bytes.  Due to the turbo code rate 1/3, this introduces a coding gain to the uncoded modulation schemes, which are, 13.75 dB for QPSK, 15.75 dB for 16-QAM, and 17 dB for 64-QAM.  Fig. \ref{fig:PDRvSINR} compares simulated PDR results to \nic{a PDR} approximation.  The PDR approximation, approximates the constants $a_c$ and $b_c$ by taking the minimum error margin between the simulated PDR and approximated PDR.  We see that this approximation result is very close to the simulation.  Table \ref{tab:pdrparam} outlines the approximated values for $a_c$ and $b_c$, for all three modulations, with the same packet length of 1024 bytes.

\subsection{User-Behavior}
User-behavior is considered in wireless communications in order to enhance user experience and \da{their perceived benefit} within the network.  Here, we assign the BS to measure satisfaction of the cell and the D2D users.  \ns{Satisfaction is measured according} to the BS's knowledge of the cell and \nss{its perceived benefit} of the D2D users.  Hence, \ns{the BS has perfect information about the current state of the cell at each stage and} \da{measures perceived benefit across all users/followers}  \ns{based on feedback from the followers}.  The BS's satisfaction is represented by $x$, and is bounded between $(0,1]$.  Satisfaction of 0 means the BS has no satisfaction with the cell and the D2D users; and satisfaction of 1 means the BS has total satisfaction with the cell and the D2D users.  We have disregarded the case where the BS has no satisfaction, as in a real world scenario the BS would have some satisfaction with the cell and \nss{perceived benefit of the D2D users}.  On the other hand, we categorize D2D user-behavior into three classes:
\begin{description}
    \item[(i)]   \nic{Casual};
    \item[(ii)]  \nic{Intermediate};
    \item[(iii)] \nic{Serious}.
\end{description}

\begin{table}[!b]
\begin{threeparttable}
  \begin{center}
  \caption{U-BeAS Behavior Class Properties}\label{tab:behavior}
  \begin{tabular}{|p{5cm}|p{5cm}|p{5cm}|}
  \hline
  \textbf{Casual-Behavior}           & \textbf{Intermediate-Behavior} & \textbf{Serious-Behavior}     \\ \hline
  SMS (Short Messaging Service)      & Video calling                  & Gaming                        \\ \hline
  MMS (Multimedia Messaging Service) & Data sharing                   & Uploading to the internet     \\ \hline
  Local voice call \tnote{1} & Local security and safety \tnote{2} & Downloading from the internet \\ \hline
  Lower data demand                  & Medium data demand             & Higher data demand            \\ \hline
  Lower transmit power               & Lower-Medium transmit power    & Higher transmit power         \\ \hline
  Lesser reliability                 & Medium reliability             & Higher reliability            \\
  \hline
  \end{tabular}
  \begin{tablenotes}
  \item[1] It is believed that in fifth generation (5G) cellular networks and future networks, voice calls will be sent via packets instead of bits, and possibly referred to as VoLTE (\emph{Voice-over-LTE})\cite{paisal2010seamless}.
  \item[2] Local security and safety communications would be utilized in a scenario where local coordination and communication between neighboring police officers, fireman, and ambulance officers would be necessary.
  \end{tablenotes}
  \end{center}
\end{threeparttable}
\end{table}

The D2D user-behavior classes were designed based on data demand, required transmit power, and reliability of communications.  Table \ref{tab:behavior} clearly outlines the resource requirements and some examples of practical applications (\cite{liu2014device},\cite{andreev2015network}) for each behavior class.  To ensure all D2D users guarantee QoE, D2D users should be grouped into the behavior classes which best describes their application and requirements.  \nic{Overall, the BS aims to maximize its satisfaction such that social welfare is guaranteed across all D2D users.  Social welfare is} \ds{guaranteed} \nic{when the leader achieves total satisfaction} \dsn{with} \nic{the whole cell and} \ds{all} \nic{D2D pairs} \ds{within the cell.}

\section{U-BeAS: User-Behavior-Aware Stackelberg Game} \label{sec:4}
\begin{figure}[!t]
        \centering
        \includegraphics[width=0.9\columnwidth]{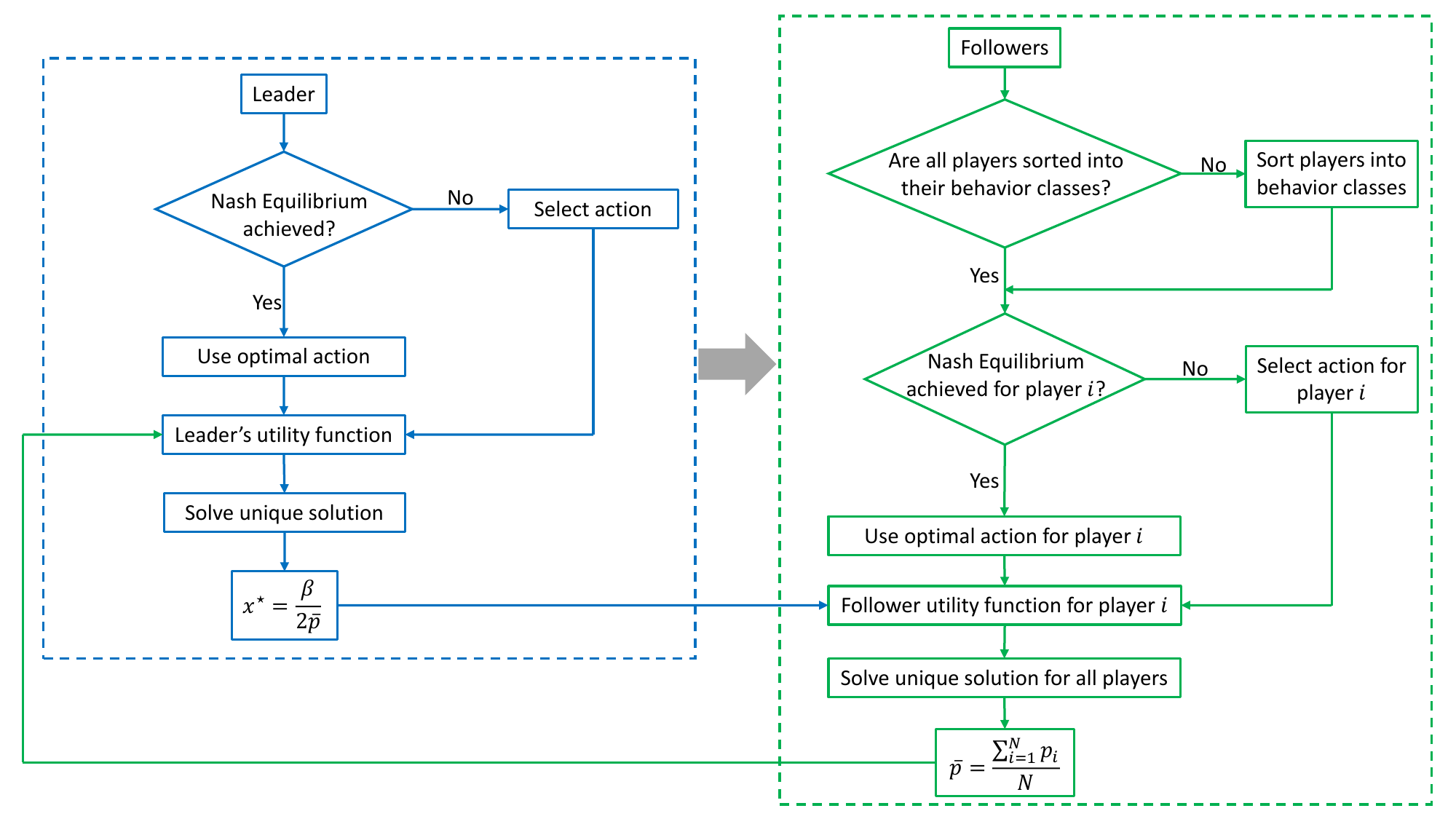}
        \caption{U-BeAS game architecture in one time step $t$.}
        \label{fig:architecture}
    \end{figure}

We propose \nic{U-BeAS, a user-behavior-aware} dynamic Stackelberg repeated game, which is finitely repeated for $T$ times, $0 \leq t \leq T$, where $T<\infty$.  The \nic{proposed} game consists of a single leader and multiple followers\ds{, where t}he leader of the game is the BS and the D2D pairs are the followers, as outlined in Fig. \ref{fig:architecture}.  The BS aims to gain optimal satisfaction \dsn{with} the cell, with respect to the D2D users.  Initially, the BS has little satisfaction and \nss{perceived benefit} about the cell and the D2D users.  Thus, the BS assigns a satisfaction price to the D2D \nic{pairs} to gain more \nss{perceived benefit about the cell and the D2D pairs}. The satisfaction price assigned to the D2D \nic{pairs} must be reasonable, such that appropriate levels of interference and transmit power are maintained by D2D \nic{pairs}.  D2D \nic{pairs} on the other hand behave selfishly with respect to each other and the BS.  The followers of the proposed game have imperfect information, which means that all followers select their action simultaneously.  The followers \nic{play a non-cooperative power control game, where the} individual strategy \nic{of each follower} is to minimize their transmit power within \nic{a given} transmit power range.  When a follower selects it's action, there are two factors which must be considered: the satisfaction price set by the BS; and the aim to maximize their own utility function.  \nic{U-BeAS} is a sequential game, which is evident in Fig. \ref{fig:architecture}, thus the leader first solves its utility function and assigns a \nic{satisfaction} price to the followers.  The followers then solve their utility function and share their measured performance to the BS, such that it can update its satisfaction \dsn{with} \nic{the cell}, for the next iteration in the game.

\nic{Initially,} D2D users \nic{sort themselves into one of the behavior classes (as in Fig. \ref{fig:architecture}), such that} QoE \nic{is guaranteed across all users.}  Each \nic{behavior class} has its own utility, due to the different requirements for each behavior, with respect to transmit power and PDR.  Once D2D users have assigned themselves to a behavior group, they cannot change behaviors during the finite stages of the repeated game, as their actions are deterministic and thus are pure strategies.  For a fair comparison between the behavior \nic{classes}, we ensure that there is an even number of D2D transmitters (from the D2D pairs) in each behavior \nic{class}.  Since we consider only one user within a D2D pair transmitting at any one time, the receiver from the D2D pair can either lie within the same behavior \nic{class} or another.

\subsection{Leader Utility Function Analysis}\label{sec:l}
The leader's utility function, $U_{BS}(x,\bar{p})$, is a function of satisfaction, $x$, and average measured performance of the followers, i.e., average transmit power, $\bar{p}$.  The utility function for the leader of the proposed game is a quadratic function, chosen for its particular and simple properties.  The action set of the leader is a finite set, where it measures its satisfaction between $(0,1]$.  \ns{The utility function designed for the leader is simply used to measure satisfaction within leader's strategy space, which enables suitable best response to the BS, while the followers are able to achieve an efficient outcome.}  \nic{After the first iteration, }the leader reacts to the average measured performance of the followers, $\bar{p}$, by updating its satisfaction with\nic{in} the cell accordingly.  The utility function for the leader (BS) is as follows,
    \begin{equation} \label{eq:UL}
        U_{BS}(x,\bar{p}) = -\bar{p} x^2 + \beta x + \kappa,
    \end{equation}
    \begin{equation*}
        \begin{aligned}
            \bar{p} &= \frac{\sum_{i=1}^{M}p_{di}(t-1)}{M}, &\\
            \beta &= 2\bar{p} x \ln(t), &\\
            \kappa &= \kappa_c\ln(t),
        \end{aligned}
    \end{equation*}
where $p_{di}(t-1)$ is the followers transmit power from the previous time slot; $t$ is the iteration/time step of each stage in the proposed game, $t\in[0,T]$; $\kappa_c$ is a positive scalar, $\kappa_c>0$.  At the end of each iteration in the proposed game, the leader finds the best response satisfaction of the cell with respect to the D2D pairs, by maximizing its utility function \eqref{eq:UL} with respect to the leader's satisfaction, given by,
    \begin{equation*}
        \begin{aligned}
            & \underset{x}{\text{maximize}}
            & & U_{BS}(x,\bar{p})~\forall~x\\
            & \text{subject to}
            & & x \in (0,1].
        \end{aligned}
    \end{equation*}

\subsection{Followers Utility Function Analysis}\label{sec:f}
The followers play a non-cooperative power control game\nic{, which} \ns{provides an optimal trade-off between} minimiz\ns{ing} transmit power \ns{and maximizing PDR with respect to D2D user-behavior} \nic{for all players}.  \ns{Minimizing transmit power for all players, will result in a reduction} \da{of radio} \nic{interference within the network}.  The action set implemented for the followers is a finite set and is bounded by a minimum transmit power and maximum transmit power, $p_{min} \leq p_{di} \leq p_{max}$.  The followers further consider QoE and initially select a behavior which best suits their \nic{resource} requirements and practical application.  The three \nic{behavior classes} considered in U-BeAS are assigned different utility functions\nic{, where each utility function is charged a satisfaction price assigned by the leader.  The satisfaction price has been designed in terms of optimal BS satisfaction and individual user transmit power.  In order to solve for optimal transmit power and PDR for all followers, the utility functions \eqref{eq:UF1}, \eqref{eq:UF2}, and \eqref{eq:UF3} from each behavior class (where $\star$ represents any behavior class), are maximized at the end of each iteration, as follows,}
    \begin{equation*}
        \begin{aligned}
            & \underset{p_{di}}{\text{maximize}}
            & & U_{F\star,i}(x,p_{di})~\forall~i~\in~\mathcal{M}\\
            & \text{subject to}
            & & [p_{min},p_{max}]mW.
        \end{aligned}
    \end{equation*}

\subsubsection{\nic{Casual-Behavior} Utility Function}\label{sec:f1}
The \nic{casual-behavior} utility function for D2D \nic{pairs}, has been taken from \cite{koskie2005nash}, $U_{FC,i}(x,p_{di})$.  This utility function is an SINR-balancing function and is modified to include the \nic{BS satisfaction price}.  This type of utility function was selected for this behavior as it gives improved emphasis on transmit power and less emphasis on PDR, while still maintaining minimum target PDR.  The practical application for this behavior requires low\nic{er} transmit power and quality compared to the other two behavior \nic{classes}.  The utility function for causal-behavior is defined as follows, 
    \begin{equation}\label{eq:UF1}
        U_{FC,i}(x,p_{di}) = \underbrace{\bigg(\frac{\bar{\gamma}_{d}}{\gamma_{di}}\bigg)p_{di}}_{\substack{\text{D2D pair}\\ \text{Performance}}} - \underbrace{D_i(x,p_{di})}_{\substack{\text{BS Satisfaction}\\ \text{Price}}},
    \end{equation}
where $p_{di}$ is transmit power for the $i$-th D2D user; $\bar{\gamma}_{d}$ is target SINR for D2D users, which is set by target PDR; $\gamma_{di}$ is SINR for the $i$-th D2D user; $D_i(x,p_{di})$ is the \nic{BS satisfaction price} which is analyzed in Section \ref{sec:f4}.

\subsubsection{\nic{Intermediate-Behavior} Utility Function}\label{sec:f2}
The \nic{intermediate-behavior} utility function is taken from \cite{koskie2005nash}, where the utility function was originally a cost function.  \nic{A cost function can be transformed into a utility function, by taking the negative of the cost function.  The cost function from \cite{koskie2005nash} is transformed into a utility function, and is further modified to include the BS satisfaction price, $U_{FI,i}(x,p_{di})$}.  We have assigned the following utility function to this behavior, as it gives \nic{improved emphasis on transmit power} and approximately equal emphasis on PDR.  The practical application for this behavior requires \nic{higher resource requirements than the casual-behavior class}, which means a slightly higher demand for transmit power, to ensure good signal quality at the receiver without larger latency periods.  The utility function for \nic{intermediate-behavior} for any player, is given by,
    \begin{equation}\label{eq:UF2}
        U_{FI,i}(x,p_{di}) =\underbrace{-sp_{di} - c\big(\bar{\gamma}_{d} - \gamma_{di} \big)^2}_{\text{D2D pair Performance}}      -\underbrace{D_i(x,p_{di})}_{\substack{\text{BS Satisfaction} \\ \text{Price}}},
    \end{equation}
where $p_{di}$ is transmit power for the $i$-th D2D user; $\gamma_{di}$ is SINR for the $i$-th D2D user; $s>0$ and $c>0$ are positive scalars; $D_i(x,p_{di})$ is the \nic{BS satisfaction price}, which is analyzed in Section \ref{sec:f4}.

\subsubsection{\nic{Serious-Behavior} Utility Function}\label{sec:f3}
The \nic{serious-behavior} utility function \nic{has been} taken from \cite{dong2015socially}, where the utility function is a function of transmit power and PDR, $U_{FS,i}(x,p_{di})$.  We \nic{have} modif\nic{ied} the utility function from \cite{dong2015socially} to include \nic{the BS satisfaction price}.  This utility function has been assigned to this behavior as it gives larger emphasis on PDR and less emphasis on transmit power.  Users implementing this behavior would require significantly higher transmit power than the other two behavior \nic{classes, for example,} uploading and downloading of large content, \nic{gaming,} and general higher throughput requirements.  Due to the greater emphasis on higher PDR, this means that signal quality is ensured, communication reliability is guaranteed, content \nic{can be} fully uploaded/downloaded in a reasonable timeframe\nic{, and serious gamers won't experience} any significant latency concerns.  The utility function for the \nic{serious-behavior class}, is given by,
    \begin{equation}\label{eq:UF3}
        U_{FS,i}(x,p_{di}) = \underbrace{-{p_{di}}^w-\frac{h_i}{{pdr_{di}}^v}}_{\text{D2D pair Performance}} - \underbrace{D_i(x,p_{di})}_{\substack{\text{BS Satisfaction} \\ \text{Price}}},
    \end{equation}
where $p_{di}$ is transmit power for the $i$-th D2D user; $pdr_{di}$ is PDR for the $i$-th D2D user; $h_i>0$ is a positive scalar; $w>0$ and $v>0$ are positive exponents; $D_i(x,p_{di})$ is the \nic{BS satisfaction price}, which is analyzed in Section \ref{sec:f4}.

\subsection{BS Satisfaction Price}\label{sec:f4}
The three \nic{follower} utility functions, \eqref{eq:UF1}, \eqref{eq:UF2}, and \eqref{eq:UF3}, are assigned a \nic{BS satisfaction price, which is also known as a cost function}.  The aim of the \nic{satisfaction price} is to charge the followers for using the dedicated spectrum \nic{and} \da{enable the BS} \nic{to attain} \da{the best perceived benefit, and social welfare, across the cell.}  The cost function we have created \nic{takes into account} both the leader's optimal satisfaction, $x$, and the individual follower's transmit power, $p_{di}$.  The \nic{BS satisfaction price} is given by,
    \begin{equation} \label{eq:cost}
        D_i(x,p_{di}) = \bigg(\frac{\Delta}{\ln(q-x)}\bigg)\bigg(\frac{1}{\ln(y-\frac{p_{di}}{z})}\bigg),
    \end{equation}
where $\Delta>0$, $q>0$, $y>0$, $z>0$ are positive constants.

Equation \eqref{eq:cost} logarithmically combines the leader's satisfaction $x$ and the follower's individual transmit power $p_{di}$.  As a result, the followers utility functions will decrease with increasing cost.  \nic{We have designed the cost function using natural logarithms as they are slowly varying functions with respect to variable changes. A} \ds{slowly varying cost function is added} \nic{to the follower's utility function} \ds{to give the most noticeable} \nic{change in utility when the leader becomes optimally satisfied with the cell, along with the followers achieving optimal transmit power.  In Appendix \ref{appendix:part1}, we prove that the BS satisfaction price is strictly concave and continuous with respect to both follower's transmit power and BS satisfaction.}

\begin{define} \label{def}
The pair of strategies $\mathbf{A^*}=[x,p_{di}]=[(x^\star,p_1^\star),(x^\star,p_2^\star),\ldots,(x^\star,p^\star_{M})]$ for the leader (BS) and all the followers (D2D pairs), $i \in \mathcal{M}$, there exists a Stackelberg Equilibrium in the U-BeAS game, if the following is satisfied:
     \begin{enumerate}
       \item The leader first achieves a Nash Equilibrium:
            \begin{equation*}
                U_{BS}([x^\star,\bar{p}]) \geq U_{BS}([x,\bar{p}])~\forall~x.
            \end{equation*}
       \item The followers then achieve a Nash Equilibrium thereafter:
            \begin{equation*}
                U_{F,i}([x^\star,p_{di}^\star],[\mathbf{x^\star},\mathbf{p^\star_{-di}}]) \geq U_{F,i}([x,p_{di}],[\mathbf{x^\star},\mathbf{p_{-di}^\star}])~~\forall~p_{di}\in\mathbf{p},~\forall~x,~\text{and}~\forall~i\in\mathcal{M}.
            \end{equation*}
     \end{enumerate}
\end{define}

\begin{remark}
Once the leader achieves a \dsn{unique} Nash Equilibrium, the followers will then achieve a \dsn{unique} Nash Equilibrium thereafter, resulting in a \dsn{unique} Stackelberg Equilibrium.
\end{remark}

\begin{proposition}
The \dsn{unique} Stackelberg Equilibrium, i.e., the best response for the leader and all the followers, exists over all stages of the finite game. Furthermore it is sub-game perfect as it is independent of the game history (i.e., previous stages in the game).
\end{proposition}

\dsn{Following from} Definition \ref{def} we \dsn{demonstrate} that there exists \nic{a} \dsn{unique} \nic{Nash Equilibrium for the leader and followers} for the proposed game, across all players, \nic{and further prove that there exists a} \dsn{unique} \nic{Stackelberg Equilibrium overall,} in the following two proofs.

\begin{IEEEproof}
\nic{Firstly, we observe that satisfaction, $x$, is a nonempty, convex, and compact subspace of Euclidean space $\mathds{R}^N$.  This condition is satisfied, as the leader's strategy space is bounded by $(0,1]$.  Next we prove that the leader's utility function \eqref{eq:UL} will have a} \dsn{unique} \nic{Nash Equilibrium at each stage in the repeated game, by solving the first order derivative over the leader's strategy space, as follows,}
    \begin{equation} \label{eq:dUBS_dx}
            \frac{\partial U_{BS}(\cdot)}{\partial x} = -2\bar{p} x +\beta.
    \end{equation}

From \eqref{eq:dUBS_dx} we can see that \eqref{eq:UL} is a continuous function within the range $(0,1]$.  \nic{For a Nash Equilibrium to have a unique solution, we set the first derivative to zero, and solve for a unique maximum solution for satisfaction, as follows,}
    \begin{equation} \label{eq:xstar}
        x^\star = \frac{\beta}{2\bar{p}}.
    \end{equation}

\nic{The unique solution of the leader's satisfaction is dependent upon both $\beta$ and $\bar{p}$.}  We input $x^\star$ \nic{directly} into the followers utility functions \nic{accordingly}.

\nic{Furthermore, we prove that} the \nic{leader's} utility function \eqref{eq:UL} is strictly concave, \nic{by solving the} second order derivative condition\nic{, i.e., $\frac{\partial^2U_{BS}}{\partial x^2}<0$},
    \begin{equation}\label{eq:d2UBS_dx2}
            \frac{\partial^2 U_{BS}(\cdot)}{\partial x^2} =-2\bar{p}~<~0,
    \end{equation}
\nic{where $\bar{p}$ always meets the condition, $\bar{p}~>~0$.  Overall, the leader's utility function \eqref{eq:UL} is strictly concave and continuous over the leader's strategy space, $(0,1]$.}
\end{IEEEproof}

\begin{IEEEproof}
\nic{More particularly, we observe that $p_{di}$ is a nonempty, convex, and compact subspace of Euclidean space $\mathds{R}^N$.  This condition is satisfied, as the followers strategy space is bounded by $[p_{min}, p_{max}]$.  We then} prove that \nic{there exists a} \dsn{unique} \nic{Nash Equilibrium at each stage in the repeated game for each of the} three utility functions for all followers, \eqref{eq:UF1}, \eqref{eq:UF2}, \eqref{eq:UF3}.  \nic{The first order derivative for each follower's utility function is solved over the follower's strategy space, $[p_{min},p_{max}]$, as follows,}
\begin{description}
    \item[(i)]   \nic{\textit{Casual-Behavior}}:\
                    \begin{equation} \label{eq:dUF1_dp}
                    \begin{aligned}
                        \frac{\partial U_{FC,i}}{\partial p_{di}} =& -A, &\\
                        \text{where} ~ A =& \frac{\Delta}{z(y-\frac{p_{di}}{z})\ln(q-x)(\ln(y-\frac{p_{di}}{z}))^2},&\\
                    \end{aligned}
                    \end{equation}
    \item[(ii)]  \nic{\textit{Intermediate-Behavior}}:\
                    \begin{equation} \label{eq:dUF2_dp}
                        \frac{\partial U_{FI,i}}{\partial p_{di}} = -s+\frac{2c\gamma_{di}(\bar{\gamma}_d-\gamma_{di})}{p_{di}} - A,
                    \end{equation}
    \item[(iii)] \nic{\textit{Serious-Behavior}}:\
                    \begin{equation} \label{eq:dUF3_dp}
                        \frac{\partial U_{FS,i}}{\partial p_{di}} = -w{p_{di}}^{w-1}+\frac{h_ivab(\gamma_{di})^b}{p_{di}(pdr_{di})^{v}} - A.
                    \end{equation}
\end{description}

From \eqref{eq:dUF1_dp}, \eqref{eq:dUF2_dp} and \eqref{eq:dUF3_dp} we can see that \nic{all of the} follower utility functions are continuous within the range $[p_{min},p_{max}]$.  \nic{For the follower utility functions to have a unique Nash Equilibrium, we set the first derivative to zero, and }substitute $x^\star$ for $x$ from \eqref{eq:xstar}.  \nic{The unique Nash Equilibrium, is a unique maximum solution} for each player's transmit power, ${p_{di}}^\star$.  \nic{The followers provide an average of all the players unique solutions, such that the leader can update its satisfaction in the next iteration, i.e., the followers solve an average unique solution}, $\bar{p}$.  Therefore, the Nash Equilibrium is a hierarchial equilibrium, as the leader first achieves a Nash Equilibrium and the followers achieve Nash Equilibrium thereafter.

\nic{Moreover, we further prove that the} followers utility functions are strictly concave, \nic{by solve the} second order derivative condition, $\frac{\partial^2U_{F\star}}{\partial {p_{di}}^2}<0$.
\begin{description}
    \item[(i)]      \nic{\textit{Casual-Behavior}:}\
                        \begin{equation} \label{eq:d2UF1_dp2}
                        \begin{aligned}
                            \frac{\partial^2 U_{FC,i}}{\partial {p_{di}}^2} =& -B~<~0, &\\
                            \text{where} ~ B =& \bigg(\frac{\Delta}{(y-\frac{p_{di}}{z})^2z^2\ln(q-x)(\ln(y-\frac{p_{di}}{z}))^2}\bigg)\bigg(\frac{2}{\ln(y-\frac{p_{di}}{z})}+1\bigg),
                        \end{aligned}
                        \end{equation}
    \item[(ii)]     \nic{\textit{Intermediate-Behavior}:}\
                        \begin{equation} \label{eq:d2UF2_dp2}
                            \frac{\partial^2 U_{FI,i}}{\partial {p_{di}}^2} = -2c\bigg(\frac{\gamma_{di}}{p_{di}}\bigg)^2 - B~<~0,
                        \end{equation}
    \item[(iii)]    \nic{\textit{Serious-Behavior}:}\
                        \begin{equation} \label{eq:d2UF3_dp2}
                            \frac{\partial^2 U_{FS,i}}{\partial {p_{di}}^2} = -w(w-1){p_{di}}^{w-2} + \frac{abh_iv(\gamma_{di})^b}{{p_{di}}^2(pdr_{di})^{v}}\bigg(-1+b-abv(\gamma_{di})^b\bigg) - B~<~0,
                        \end{equation}
\end{description}

\nic{For casual-behavior utility \eqref{eq:d2UF1_dp2}} to satisfy the second order derivative condition $B$ must be greater than zero $(B>0)$\nic{, hence the following variables must satisfy these conditions;} $\Delta>0$, $q>0$, $y>0$, and $z>0$\nic{, and} we know that $x$ and $p_{di}$ are always positive.  \nic{For the intermediate-behavior utility \eqref{eq:d2UF2_dp2} to satisfy the concave condition,} $2c\big(\frac{\gamma_{di}}{p_{di}}\big)^2>0$, $c>0$, and $B>0$\nic{, and we know that} $\gamma_{di}$ is always positive.  \nic{Finally, for the serious-behavior utility \eqref{eq:d2UF3_dp2} to satisfy the second order derivative condition,} we split \eqref{eq:d2UF3_dp2} into three sections.  \nic{The first and third sections will always be negative, as $w\geq 1$, and all conditions set on variables within $B$ are satisfied.  In order for the} second part of \eqref{eq:d2UF3_dp2} to satisfy the \nic{concavity condition,} the \nic{following} constants must satisfy these conditions; $a<0$, $b<0$, $h_i>0$, and $v>0$\nic{, and we know that} $pdr_{di}$ is always positive.  In appendix \ref{appendix:part1}, we prove that each of the follower's utility functions \eqref{eq:UF1}, \eqref{eq:UF2}, and \eqref{eq:UF3} are strictly concave and continuous with respect to the leader's satisfaction.
\end{IEEEproof}

\begin{IEEEproof}
\nic{A unique Nash Equilibrium exists across all followers, even \ds{though} their utility functions are different.  We have individually proved that the three utility functions used by the followers are strictly concave, continuous, and compact, across the same strategy space for all players, i.e., $[p_{min},p_{max}]$. Following from the proof in \cite{rosen1965}, if all followers utility functions, whatever they may be, are strictly concave and continuous in the same strategy space that is compact, then the unique equilibrium follows. 
}
\end{IEEEproof}

\begin{define}
The game is Pareto-efficient when the leader's satisfaction is maximized, and all the followers transmit powers are minimized, while achieving desired PDR in the U-BeAS game.
\end{define}

\begin{remark}
For the game to have a Pareto-efficient outcome, the leader must first converge to an optimal outcome, then the followers will converge to an Pareto-optimal outcome thereafter.
\end{remark}

\begin{proposition}
When the leader has maximized its satisfaction with the cell, which includes knowledge of an efficient outcome from the followers, it is an optimal outcome for the leader that guarantees social welfare across all the followers in the U-BeAS game.
\end{proposition}

\begin{IEEEproof}
A Pareto-efficient outcome is fully defined over \nic{all stages in the repeated} game.  The leader first obtains an efficient outcome with respect to satisfaction, then the followers converge to a Pareto-efficient outcome with respect to transmit power and PDR, such that the game achieves an overall Pareto-optimal outcome.  When the BS's satisfaction, $x$, is maximized the desired outcome in terms of satisfaction is achieved.  Hence, each D2D pair transmitter, $i$, will minimize its transmit power with respect to all other transmitters, $\mathbf{-i}$, until no D2D pair transmitter can further decrease its transmit power over $T$ stages, and at the same time desired PDR is achieved for each D2D pair.
\end{IEEEproof}

\section{Simulation \nic{Set-up and} Results} \label{sec:5}
In this section, we outline the simulation \nic{set-up and} results of our proposed game, U-BeAS.  We compare the three \nic{behavior classes} and investigate the trade-off between transmit power and PDR \nic{for all D2D pairs}.  We also look at the effect of the leader's satisfaction price, on D2D pairs.  The results of the \nic{U-BeAS game} are compared to three \nic{different} non-cooperative power control \nic{(NPC)} games.

\subsection{Simulation Set-Up}
    \begin{table} [!b]
        \begin{center}
        \caption{Simulation Parameters and Values} \label{tab:simulation}
        \begin{tabular}{|l|l||l|l|}
        \hline
        \textbf{Parameter}                    & \textbf{Value}          & \textbf{Parameter}                    & \textbf{Value}\\ \hline
        Cell Radius (R)                       & 500~m                   & $A_{PL}$                              & $10^{-3.22}$ \\ \hline
        Maximum Distance between D2D Pairs    & 50~m                    & Doppler Spread ($f_DT_s$)             & 0.01 \\ \hline
        Carrier Frequency                     & 2~GHz                   & Path Loss Exponent for D2D users $\varphi$          & 4 \\ \hline
        BS Per Cell                           & 1                       & D2D Transmit Power Range              & $0$-$23$~dBm \\ \hline
        D2D Pairs Per Cell $\mathcal{M}$      & 24                      & Cellular User Constant Transmit Power & 14~dBm \\ \hline
        Reference D2D Distance ($d_0$)        & 20~m                    & D2D User Noise Power ($N_0$)          & $-99.21$~dBm \\
        \hline
        \end{tabular}
        \end{center}
    \end{table}

\nic{U-BeAS} was implemented over, $T=100$, stages and repeated, $RPT=\nic{1000}$, times \ds{for a time-selective slow, and flat, Rayleigh fading channel}.  The BS is positioned at the center of the cell, and the D2D users and cellular user are placed randomly within the cell.  The distance between users in a D2D pair are random and do not  exceed the maximum D2D transmission range.  We set a minimum target PDR of 0.9 for all levels of behavior, which results in a target SINR of $0.534$, using the modulation scheme 16-QAM with 1/3 rate turbo code.  All simulation values are outlined in Table \ref{tab:simulation}.  \nic{The three different NPC games are implemented in the same scenario/set-up as the U-BeAS game.  We have compared one NPC game to each behavior class, and the utility functions have been taken from \cite{koskie2005nash} and \cite{dong2015socially}.  The NPC game procedure occurs as follows,}
\begin{enumerate}
  \item Players select an action from the action set, i.e., transmit power;
  \item \nss{Each player calculates their} utility function;
  \item \nss{The utility function is then maximized} with respect to transmit power for each player, in order to solve for optim\nss{al} transmit power;
  \item Repeat steps 1) to 3) until, $t=T$ (finite game).
\end{enumerate}

\subsection{Game Results}
\begin{figure} [!b]
        \centering
        \includegraphics[width=0.56\columnwidth]{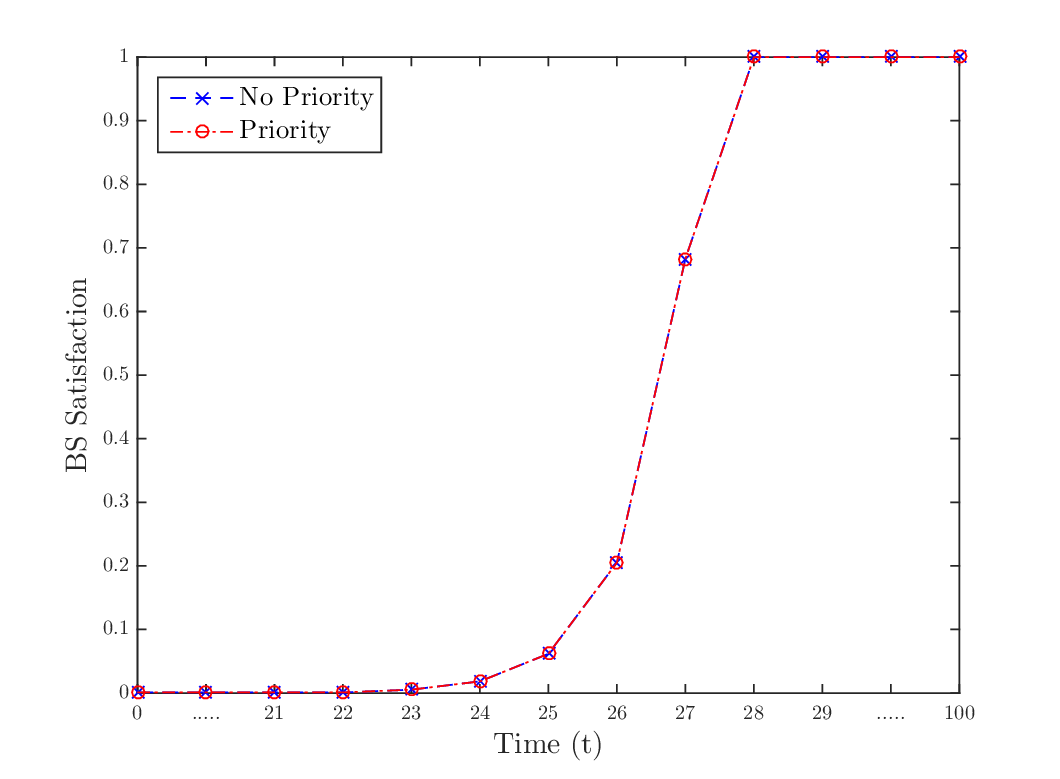}
        \caption{Leader's satisfaction evolving throughout U-BeAS, with and without priority on each behavior class.  The leader's utility function \eqref{eq:UL} constant: $\kappa_c = 4$.}
        \label{fig:Result1}
    \end{figure}

\nic{Firstly}, we look at the leader's satisfaction \dsn{with} the cell, with respect to whether the D2D pairs consider priority or not, \nic{see} Fig. \ref{fig:Result1}.  \nic{The U-BeAS game, i.e., a user-behavior-aware} dynamic Stackelberg game, maximizes the BS's measured performance and obtains optimum satisfaction.  As the game evolves over the number of stages in the game, so does the satisfaction of the BS, which grows from 0.001 to 1 during 6 stages.  The BS's satisfaction evolves due to the \nic{satisfaction price charged to the followers, and the} average measured performance \nic{of the followers} fed back to the leader at the end of each iteration of the game.  Fig. \ref{fig:Result1} shows that by \nic{prioritizing or not prioritizing each behavior class}, there is no effect on the leader's satisfaction.  Fig. \ref{fig:Result1} also shows fast convergence to an optimal satisfaction for the leader in the U-BeAS game for both cases, i.e., \nic{the BS achieves total satisfaction within the cell, $x=1$}.

    \begin{figure}[!t]
        \centering
        \subfloat[Average transmit power of the followers, without priority.]{\includegraphics[width=0.5\columnwidth]{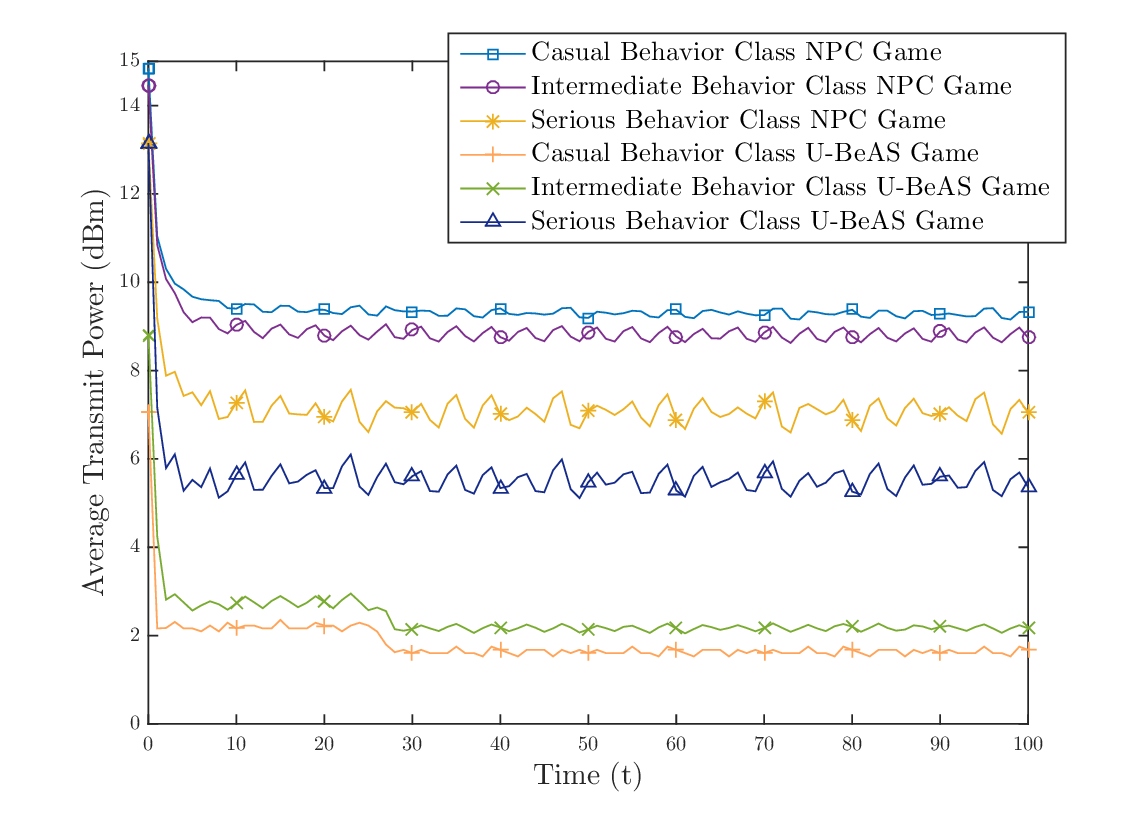}\label{fig:Result2}}
        \hfill
        \subfloat[Average PDR of the followers, without Priority.]{\includegraphics[width=0.5\columnwidth]{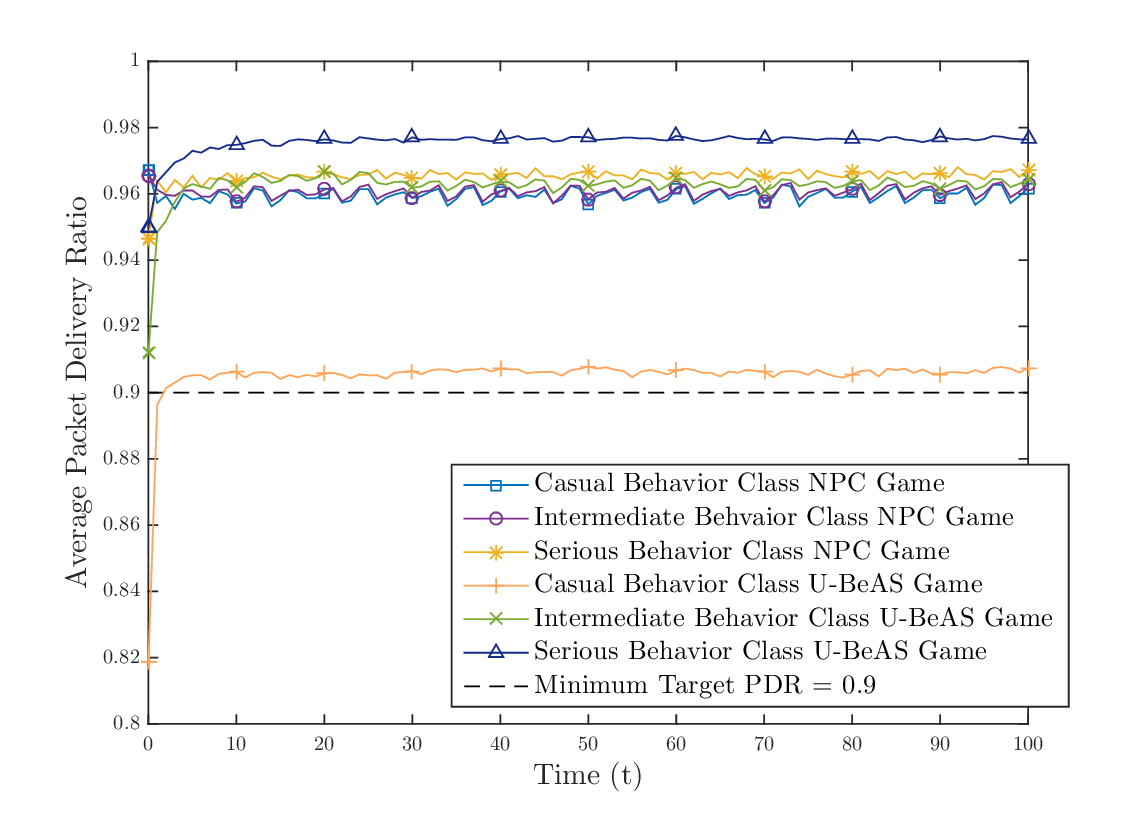}\label{fig:Result3}}
        \caption{Measured Performance of the Followers without priority.  The follower's utility functions \eqref{eq:UF1},\eqref{eq:UF2},\eqref{eq:UF3} constants are: $\Delta=1.8$, $q=3$, $y=2.001$, $z=0.6$, $s=0.05$, $c=1$, $w=2$, $v=4$.} \label{fig.R23}
    \end{figure}

     \begin{figure}[!t]
        \centering
        \subfloat[Average transmit power of the followers, with priority.]{\includegraphics[width=0.5\columnwidth]{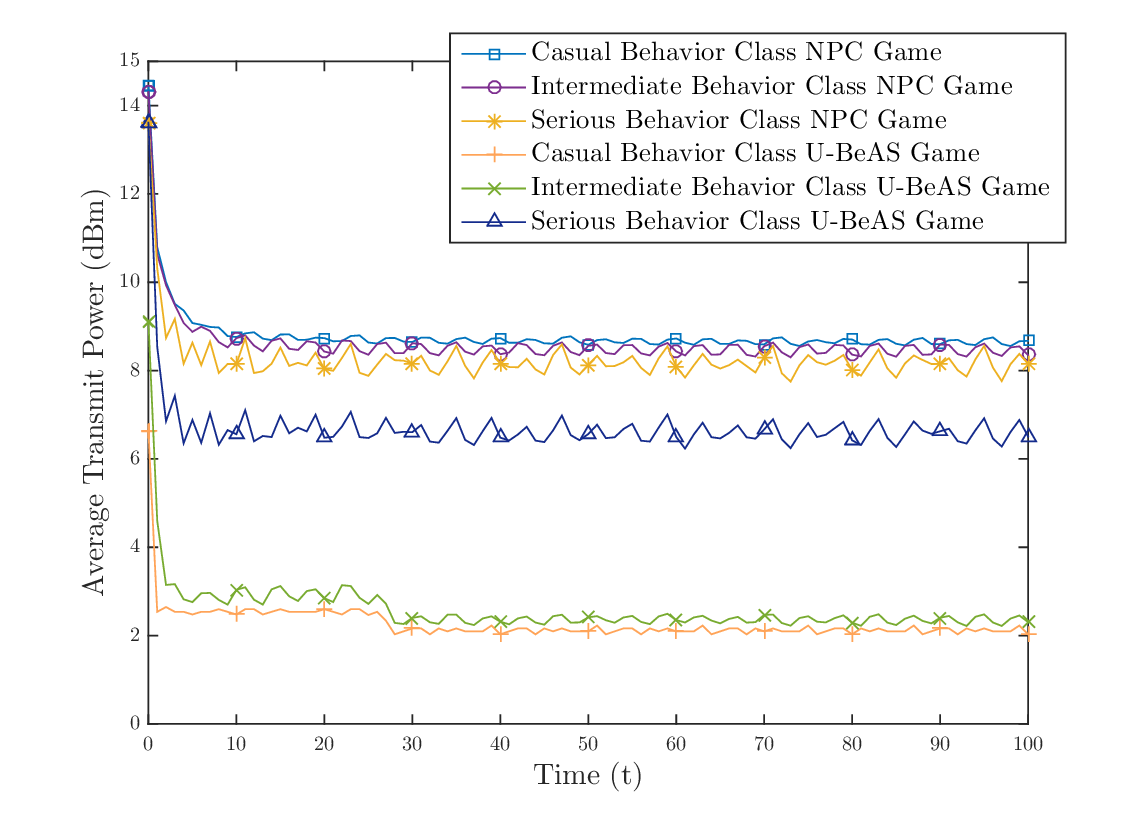}\label{fig:Result4}}
        \hfill
        \subfloat[Average PDR of the followers, with Priority.]{\includegraphics[width=0.5\columnwidth]{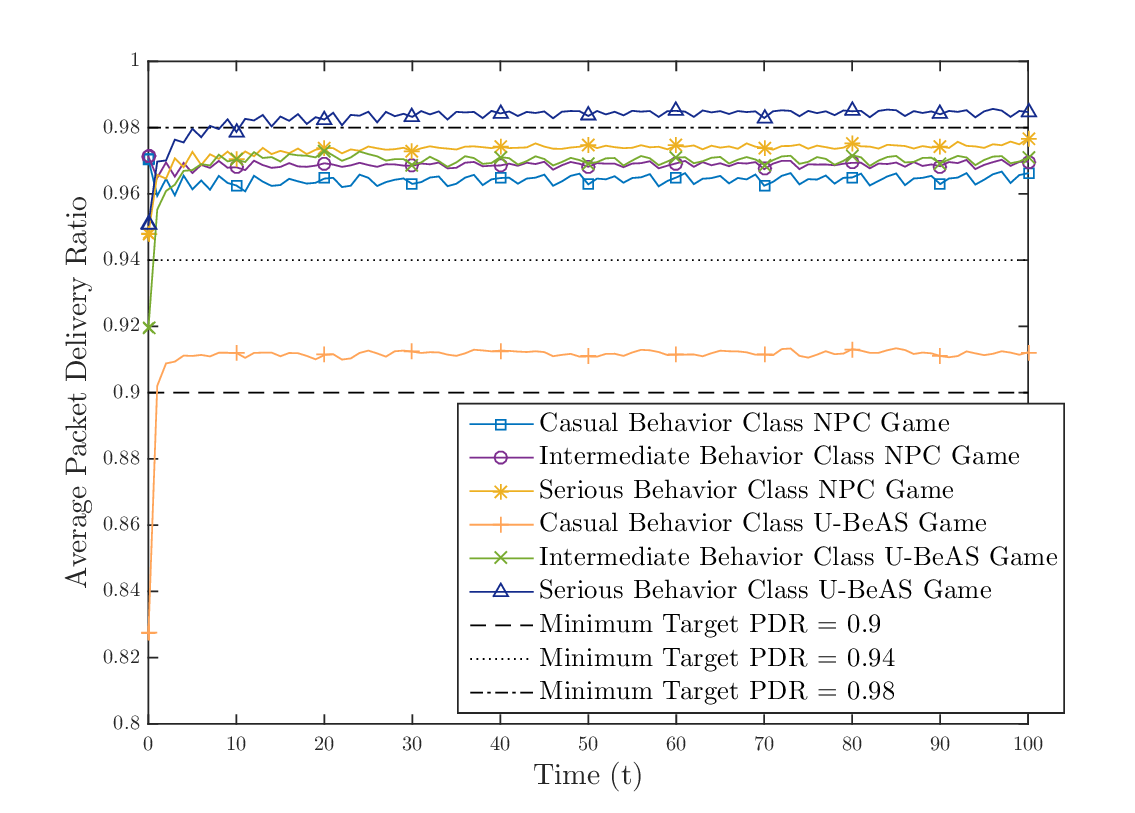}\label{fig:Result5}}
        \caption{Measured Performance of the Followers with priority.  The follower's utility functions \eqref{eq:UF1},\eqref{eq:UF2},\eqref{eq:UF3} constants are: $\Delta=1.8$, $q=3$, $y=2.001$, $z=0.6$, $s=0.05$, $c=1$, $w=2$, $v=4$.} \label{fig:R45}
    \end{figure}

\nic{Fig. \ref{fig.R23} plots the D2D pair's average performance without considering priority for each behavior class.  Average transmit power for the three behavior classes for all followers is illustrated in} Fig. \ref{fig:Result2}.  It is evident that the \nic{U-BeAS} game reduces transmit power on average more than the NPC (non-cooperative power control) games.  \nic{From Fig. \ref{fig:Result2} and Fig. \ref{fig:Result3}, the plot lines do not converge to a single value in most cases due to the effects of slow fading.}  Table \ref{tab:txresults} outlines the average outcome of transmit power for the three \nic{behavior classes,} for both the U-BeAS game, and the three NPC games.  \nic{Within} Table \ref{tab:txresults}, we include the average outcome of transmit power before and after the BS converges to an optimal outcome, for \nic{the U-BeAS} game.  As a result, once the BS converges to an optimal satisfaction, as in Fig. \ref{fig:Result1}, the followers then further reduce their transmit power to a Pareto-efficient outcome, as in Fig. \ref{fig:Result2}, while maintaining their PDR, and \nic{thus} guaranteeing social welfare across all D2D pairs.  Due to the D2D pairs within our U-BeAS game reducing their transmit power more than the three NPC games, the interference generated by D2D pairs is \ds{then} reduced significantly.  \nic{Hence,} another benefit for reducing transmit power for D2D pairs allows for an increase in battery lifetime \cite{wang2015energy}, and creating a power efficient network.
\begin{table} [!b]
        \begin{center}
        \caption{Average Transmit Power Results} \label{tab:txresults}
        \begin{tabular}{|l|l|l|l||l|l|l|}
        \hline
        \multirow{2}{*}{\textbf{Game Type}} & \multicolumn{3}{c||}{\textbf{Behavior Class without Priority}} & \multicolumn{3}{c|}{\textbf{Behavior Class with Priority}}   \\ \hhline{~------}
                                                 & \emph{Casual} & \emph{Intermediate} & \emph{Serious} & \emph{Casual} & \emph{Intermediate} & \emph{Serious}      \\ \hline
        NPC Game                                 & \nic{9.31}~dBm & \nic{8.84}~dBm & \nic{7.07}~dBm & 8.68~dBm & 8.49~dBm & 8.18~dBm     \\ \hline
        U-BeAS Game \emph{before} BS convergence & \nic{2.23}~dBm & \nic{2.73}~dBm & \nic{5.52}~dBm & 2.54~dBm & 2.91~dBm & 6.71~dBm     \\ \hline
        U-BeAS Game \emph{after} BS convergence  & \nic{1.64}~dBm & \nic{2.16}~dBm & \nic{5.55}~dBm & 2.13~dBm & 2.35~dBm & 6.62~dBm     \\
        \hline
        \end{tabular}
        \end{center}
    \end{table}

    \begin{table} [!b]
        \begin{center}
        \caption{Average Packet Delivery Ratio Results} \label{tab:pdrresults}
        \begin{tabular}{|l|l|l||l|l|}
        \hline
        \multirow{2}{*}{\textbf{Behavior Class}} & \multicolumn{2}{c||}{\textbf{Behavior Class without Priority}}  & \multicolumn{2}{c|}{\textbf{Behavior Class with Priority}} \\ \hhline{~----}
                                & \emph{U-BeAS Game}  & \emph{NPC Games}    & \emph{U-BeAS Game} & \emph{NPC Games}\\ \hline
        \emph{Casual}           & \nic{0.906}         & \nic{0.959}         & 0.912              & 0.964           \\ \hline
        \emph{Intermediate}     & \nic{0.963}         & \nic{0.960}         & 0.971              & 0.969           \\ \hline
        \emph{Serious}          & \nic{0.977}         & \nic{0.966}         & 0.983              & 0.974           \\
        \hline
        \end{tabular}
        \end{center}
    \end{table}

Fig. \ref{fig:Result3}, plots average PDR for the three behavior \nic{classes} for all followers in the \nic{U-BeAS} game, without considering communication priority.  Table \ref{tab:pdrresults} outlines the average outcome of PDR for the \nic{U-BeAS game and the three} NPC games.  We observe that the U-BeAS game has different effects on PDR, depending on the behavior class.  \nic{In Fig. \ref{fig:Result3}, D2D pairs in the U-BeAS game converge to a Pareto-efficient outcome, where D2D pairs who selected the serious-behavior class maximized PDR more than all other behavior classes within U-BeAS and all NPC games.  As a result, D2D pairs in the U-BeAS game whose behavior-class is serious will have improved communication reliability, and higher rate with little latency.  When comparing the intermediate-behavior between the U-BeAS game and the NPC game, there is little change in PDR, and as a result communication reliability is maintained.  D2D pairs in the U-BeAS game whose behavior-class is casual has the lowest PDR out of all simulation results.  However, the U-BeAS casual-behavior outcome for average PDR is still above minimum target PDR, thus maintaining} \ds{requisite} \nic{communication reliability for all users with this behavior class.}

\nic{Fig. \ref{fig:R45} plots the D2D pairs average performance considering priority for each behavior class.  By prioritizing behavior classes, this ensures particular communications will take priority over other communications, i.e., priority has been assigned to each behavior class as follows; casual-behavior has a minimum target PDR of 0.9, intermediate-behavior has a minimum target PDR of 0.94, and serious-behavior has a minimum target PDR of 0.98.}  Hence, Fig. \ref{fig:Result4} and Fig. \ref{fig:Result5} plot average transmit power and PDR for D2D pairs respectively, by taking into account different levels of priority for each behavior.  \nic{The plot lines in Fig. \ref{fig:Result4} and Fig. \ref{fig:Result5} do not converge to a steady value, which is due to the effects of slow fading.}  Fig. \ref{fig:Result4} exhibits the same characteristics as in Fig. \ref{fig:Result2}, and Table \ref{tab:txresults} outlines the average outcome of transmit power for the three behaviour classes with priority, for both the U-BeAS game and the three NPC games.  Fig. \ref{fig:Result5} on the other hand shows the effects of PDR when different levels of priority are considered for each behavior class.  Table \ref{tab:pdrresults} outlines the average outcome of PDR for the U-BeAS game and the three NPC games with priority.  Here we see that when prioritized behavior classes are considered, average PDR is slightly improved for both the U-BeAS game and the NPC games.  In Fig. \ref{fig:Result5}, the serious-behavior class for the \nic{U-BeAS} game achieves its minimum target PDR of 0.98, except this is not the case for the serious-behavior NPC game.  \nic{As a result, D2D pairs in the U-BeAS game whose behavior-class is serious will have improved communication reliability, priority, higher rate, and reduced latency.}  For the other behavior classes, the U-BeAS game and the NPC games also achieve their minimum target PDRs.

\nic{From Fig. \ref{fig:Result1},} the leader becomes more satisfied \dsn{from} the game and converges to its optimal satisfaction outcome, where the D2D pairs also further reduce their transmit power to a Pareto-efficient outcome, as in both Fig. \ref{fig:Result2} and Fig. \ref{fig:Result4}, while maintaining their PDR and converging to a Pareto-efficient outcome, as in Fig. \ref{fig:Result3} and Fig. \ref{fig:Result5}.  Therefore, once the leader converges to optimal satisfaction, the social welfare is guaranteed across all D2D users.

\section{Conclusion} \label{sec:6}
In this paper, we proposed \nic{U-BeAS, a user-behavior-aware dynamic Stackelberg repeated game.  U-BeAS} provides an optimal trade-off between minimizing transmit power and maximizing PDR, with respect to D2D user-behavior for all D2D pairs.  The BS \ds{is} the \nic{Stackelberg} leader, whose strategy was to measure satisfaction \dsn{with} the cell and the D2D users.  The D2D users were the \nic{Stackelberg} followers, whose strategy was to minimize transmit power and maintain acceptable PDR, with respect to \dsn{three possible} D2D user-behavior\dsn{s: casual; intermediate; and serious}.  Analysis of U-BeAS \dsn{proved that it has} a unique Nash Equilibrium \dsn{for both leader and followers}, thus resulting in a \dsn{unique} Stackelberg Equilibrium, which is sub-game perfect (i.e., independent of game history).  Simulation results illustrate that as the BS converges to an optimal satisfaction (total satisfaction), the followers then converge to Pareto-efficient transmit power and PDR\dsn{, while guaranteeing social welfare and quality-of-experience (QoE) across all D2D pairs}.  We compared U-BeAS to three different NPC games \ds{that} reflect the three different behaviors.  From simulation results, it is evident that U-BeAS has a much better outcome than traditional NPC games, as players transmit power is further minimized, and PDR is further maximized, with respect to D2D user-behavior for all D2D pairs.

\appendices
\section{Proof the BS Satisfaction Price is Strictly Concave and Continuous} \label{appendix:part1}
We prove that the \nic{BS satisfaction price} \eqref{eq:cost} is strictly concave and continuous with respect to the leader's satisfaction\nic{, $x$, and the follower's transmit power, $p_{di}$.  Firstly, we observe that both $x$ and $p_{di}$ are nonempty, convex, and compact subspace of Euclidean space $\mathds{R}^N$.  Next we solve the first derivative of the BS satisfaction price, firstly with respect to the leader's strategy space, and secondly to the follower's strategy space, as follows,}
    \begin{equation}\label{eq:app1}
            \frac{\partial D_i(x,p_{di})}{\partial x} = \frac{\Delta}{(q-x)(\ln(q-x))^2\ln(y-\frac{p_{di}}{z})},
    \end{equation}
    \begin{equation}\label{eq:app2}
            \frac{\partial D_i(x,p_{di})}{\partial p_{di}} = \frac{\Delta}{z(y-\frac{p_{di}}{z})\ln(q-x)(\ln(y-\frac{p_{di}}{z}))^2}.
    \end{equation}

\nic{From \eqref{eq:app1} and \eqref{eq:app2}, we see that \eqref{eq:cost} is a continuous function, both within the range $(0,1]$, and $[p_{min},p_{max}]$ respectively.  Next, to show that the leader's satisfaction price is strictly concave, we solve the second order derivative condition, $\frac{\partial^2 D_i}{\partial x^2}<0$ and $\frac{\partial^2 D_i}{\partial {p_{di}}^2}<0$, as follows,}
    \begin{equation} \label{eq:app3}
            \frac{\partial^2 D_i(x,p_{di})}{\partial x^2} = \bigg(\frac{\Delta}{(q-x)^2(\ln(q-x))^2\ln(y-\frac{p_{di}}{z})}\bigg)\bigg(\frac{2}{\ln(q-x)}+1\bigg),
    \end{equation}
    \begin{equation}\label{eq:app4}
            \frac{\partial^2 D_i(x,p_{di})}{\partial {p_{di}}^2} = \bigg(\frac{\Delta}{z^2(y-\frac{p_{di}}{z})^2\ln(q-x)(\ln(y-\frac{p_{di}}{z}))^2}\bigg)\bigg(\frac{2}{\ln(y-\frac{p_{di}}{z})}+1\bigg).
    \end{equation}

\nic{The restrictions on the variables,} $q, x, y, \Delta, z$, must satisfy their conditions as outlined in Section \ref{sec:f}.  \nic{Since, \eqref{eq:app3} and \eqref{eq:app4} are always positive, this means that when the BS satisfaction price is assigned to the followers utility it must be subtracted.  Hence, by having a negative sign in front of the BS satisfaction price, this will cause the first and second derivatives of the cost function to become negative, and thus satisfying the concavity constraint.}
\section*{Acknowledgment}
The authors would like to thank Dr Xiangyun (Sean) Zhou of the Australian National University for his helpful suggestions and comments.

\ifCLASSOPTIONcaptionsoff
  \newpage
\fi


\bibliographystyle{IEEEtran}

%




\end{document}